\documentclass[twocolumn,floatfix,superscriptaddress,amsmath,showpacs,showkeys,aps,pre]{revtex4-2}

\usepackage{epsfig}
\usepackage{bm}
\usepackage[dvipsnames]{xcolor}
\usepackage{amsmath}
\usepackage{amssymb}
\usepackage{graphicx}
\usepackage{physics}
\usepackage{xr}
\usepackage{soul}
\usepackage{longtable}
\usepackage{verbatim}
\usepackage{lmodern}
\usepackage[T1]{fontenc}
\usepackage[normalem]{ulem}
\usepackage{blkarray}

\begin{document}



\title{Susceptibility for extremely low external fluctuations and  critical behaviour of Greenberg-Hastings  neuronal model}

\author{Joaquin Almeira}
\affiliation{Instituto de F\'isica Enrique Gaviola (IFEG-CONICET) Ciudad Universitaria, 5000 C\'ordoba, Argentina.}

\author{Daniel A. Martin}
\affiliation{Instituto de Ciencias F\'isicas (ICIFI-CONICET), Center for Complex Systems and Brain Sciences (CEMSC3), Escuela de Ciencia y Tecnolog\'ia, Universidad Nacional de Gral. San Mart\'in, Campus Miguelete, San Mart\'in, Buenos Aires, Argentina.}
\affiliation{Consejo Nacional de Investigaciones Cient\'{\i}fcas y T\'ecnicas (CONICET), Buenos Aires, Argentina.}

\author{Dante R. Chialvo}
\affiliation{Instituto de Ciencias F\'isicas (ICIFI-CONICET), Center for Complex Systems and Brain Sciences (CEMSC3), Escuela de Ciencia y Tecnolog\'ia, Universidad Nacional de Gral. San Mart\'in, Campus Miguelete, San Mart\'in, Buenos Aires, Argentina.}
\affiliation{Consejo Nacional de Investigaciones Cient\'{\i}fcas y T\'ecnicas (CONICET), Buenos Aires, Argentina.}
\affiliation{Department of Physics, Faculty of Science, Hong Kong Baptist University, Hong Kong SAR, China.}

\author{Sergio A. Cannas}
\affiliation{Instituto de F\'isica Enrique Gaviola (IFEG-CONICET) Ciudad Universitaria, 5000 C\'ordoba, Argentina.}

\affiliation{Consejo Nacional de Investigaciones Cient\'{\i}fcas y T\'ecnicas (CONICET), Buenos Aires, Argentina.}
\affiliation{Facultad de Matem\'atica Astronom\'ia F\'isica y Computaci\'on, Universidad Nacional de C\'ordoba.}
\date{\today}

\begin{abstract}

We consider the scaling behaviour of the fluctuation susceptibility associated with the average activation in the  Greenberg-Hastings neural network model and its relation to microscopic spontaneous activation. We found that, as the spontaneous activation probability tends to zero, a clear finite size scaling behaviour in the susceptibility emerges, characterized by critical exponents which follow already known scaling laws. This shows that the spontaneous activation probability plays the role of an external field conjugated to the order parameter of the dynamical activation transition. 
The roles of different kinds of activation mechanisms around the different dynamical phase transitions exhibited by the model are characterized numerically and using a mean field approximation. However, while mean field approximations predict Directed Percolation critical exponents, numerical simulations on Watts-Strogatz networks provide critical exponents estimations which do not seem to agree with any known universality class.
\end{abstract}

\maketitle

\section{Introduction}
The theory of phase transitions provides relations among different observables, such as magnetization, magnetic susceptibility, temperature, and others, valid for thermodynamic equilibrium transitions. About continuous transitions, the behaviour of such variables is properly described by critical exponents,  scaling laws, and scaling relations that are ultimately based on well grounded theoretical results.
In out-of-equilibrium transitions, different quantities are frequently defined in analogy with the equilibrium observables. However, it cannot be taken for granted that the same relations will hold. In this article, we study numerically and analytically the behaviour of the susceptibility in a simple neuronal network model about a dynamical transition and its relation to an external stimulus parameter. We characterize the scaling behaviour and verify that equilibrium statistical mechanics scaling laws are followed.

The Greenberg-Hastings (GH) cellular automaton is one of the simplest neural network models displaying cooperative emergent properties \cite{Haimovici}. Adopting the G\&H formalism has been well received to model large-scale brain dynamics since the
work of Haimovici et al \cite{Haimovici} a decade ago, which successfully replicated several key observations in human brain imaging. Its simplicity allowed it to focus the attention on the role of the critical dynamics in reproducing the large-scale organization of the spatio-temporal patterns observed in the human brain under resting conditions. Since then, scores of papers have used the same approach to study the brain dynamics under different conditions as sleep, anaesthesia, coma, stroke, influence of psychedelics, etc.. Despite its simplicity, the GH neural model exhibits a rich dynamical behaviour, at least as far as criticality is concerned, including network topology dependent first and second order phase transitions \cite{Zarepour}-\cite{Almeira2024}, as well as tricritical behaviour when both inhibitory and excitatory units are considered \cite{Almeira2022}. A central quantity in all these phenomena is the activation threshold $T$ of the artificial neurons, which acts as a control parameter for all the phase transitions the system undergoes for certain network topologies (in the case of purely excitatory units, it is the only control parameter). Then, two different (although closely related) quantities exhibit the typical behaviour of an order parameter as $T$ changes: the average fraction of neurons in the largest cluster of excited neurons (thus describing a percolation-like phase transition \cite{Haimovici,Zarepour,Grigera}) and the average {\it total} fraction of excited neurons \cite{Sanchez,Grigera}. Both quantities become negligible above a certain characteristic (i.e., critical) value of $T$ and take large values below it. The curious fact is that, at least for some network topologies, both phenomena seem to happen at {\it different} critical values of the control parameter $T$, even in the infinite size limit \cite{Grigera}. Moreover, the susceptibilities associated with both order parameters appear to exhibit different scaling properties: while the average cluster size (the standard susceptibility in percolation phenomena) exhibits a size-dependent peak with associated critical exponents in the standard mean field percolation universality class \cite{Zarepour}, the fluctuation susceptibility associated with the total activity seems to exhibit a size-independent peak  \cite{Grigera}. Understanding the last fact is expected to give an important clue to the general comprehension of this complex critical scenario. The main part of the present work focuses on the finite size susceptibility problem.

Two possibilities appear to be the most probable explanations for the lack of a size-divergent peak in the activity susceptibility. On one hand, the model contains a constant (small) spontaneous activation probability $r_1$, which prevents the system from falling into an absorbing state \cite{Haimovici}.  This parameter may act as an external field conjugated to the total activity order parameter (such a kind of effect was observed in the Kinouchi-Copelli neural network model \cite{Kinouchi}), thus driving the system out of criticality and therefore eliminating the finite size scaling.  On the other hand, the GH neural model has random synaptic weights, which act as a source of quenched disorder, besides other sources of disorder, such as the intrinsic randomness of small world networks \cite{Zarepour}. It is known that quenched disorder can modify or even suppress second-order phase transitions 
\cite{Vojta2006,Vojta2019}.

To elucidate the relevance of the spontaneous activation, we analysed the critical behaviour of the GH model defined on a Watts-Strogatz network without spontaneous activation probability $r_1=0$. The numerical analysis in this case requires taking into account all the usual subtleties of absorbing phase transitions \cite{Marro,Henkel}. We applied then three different methods to obtain the behaviour of both the order parameter (total activity) and its associated susceptibility. We observed that, when removing the spontaneous activity $r_1$, a clear critical behaviour emerges, despite the presence of disorder.  Moreover, the usage of different algorithms allowed us to obtain a consistent estimation of the associated critical exponents that do not agree (to the best of our knowledge) with any previously known universality class.

To get a better understanding of the influence of the spontaneous activation on the dynamics of the model, we also analysed whether the activation mechanisms triggered by it become relevant to the order-disorder phase transition, that is, in the neighbourhood of the critical point. We found that, for low connectivity networks, single activation mechanisms triggered by the spontaneous activity make a substantial contribution to the order parameter close to the critical point. Such contribution decreases as the degree of the network increases,  the activation transition being progressively dominated by multiple neurons' cooperative mechanisms.

The paper is organised as follows. In section \ref{model} we present the GH model and different statistical quantities to be analysed in the paper. In section \ref{r10} we analyse the critical properties of the model in the absence of spontaneous activation, while in section \ref{activation} we analyse the influence of the spontaneous activation probability on the activation mechanism close to the critical point. Some conclusions are discussed in section \ref{Discussion}.


\section{The  Model}
\label{model}

We considered a neural model first introduced by Haimovici {\it et al.} \cite{Haimovici} on a network derived from the human connectome \cite{Hagmann} and later implemented in a small-world network by Zarepour {\it et al.} \cite{Zarepour}. The dynamics is that of a Greenberg-Hastings cellular automaton \cite{GH}, where each site $i$ of a network has associated a three-state variable $x_i = 0,\,1,\,2$, representing the following dynamical states: quiescent ($x_i=0$), excited ($x_i=1$) and refractory ($x_i=2$). The network is constructed from a one-dimensional ring with homogeneous degree $k$. Following the usual Watts-Strogatz (WS) recipe \cite{WS}, bonds are rewired with probability $\pi$, so the average degree $\langle k \rangle$ remains constant. 
The model follows a parallel dynamics in a discrete-time $t$ as follows: (a) each quiescent neuron can become excited with a small spontaneous activation probability $r_1$, or if the sum of their active neighbours, weighted through the connectivity matrix, surpasses an activation threshold $T$ (the control parameter of this model), which is the same for all neurons; (b) active neurons become refractory after one step, and (c) refractory neurons become quiescent with probability $r_2$ per unit time (so, the average refractory period has a duration of $1/r_2$). The transition probabilities  for the \textit{i}-th site are given by

\begin{eqnarray}\label{eq:probGH}
    P_i(0\rightarrow1) &=& 
    1-(1-r_1) \times\nonumber \\
    &&\times \qty[1-\Theta\qty(\sum_{j} W_{ij} \delta(x_j,1)-T)]\nonumber \\
    P_i(1\rightarrow2) &=& 1\nonumber\\
    P_i(2\rightarrow0) &=& r_2
\end{eqnarray}

\noindent where $\Theta$ is the Heaviside function and $\delta(x,y)$ is the Kronecker delta function. $W_{ij}$ is a symmetric synaptic matrix, and the sum is carried out over the nearest neighbours of the node $i$. Notice that  $P_i(0\rightarrow1)$ is written as $1$ minus the probability of not being activated, which is the product of probabilities of not having a spontaneous activation, $(1-r_1)$, and not having an activation due to active neighbours. Attempting to mimic the weight distribution of the human connectome \cite{Hagmann}, synaptic weights are quenched variables chosen from an exponential distribution

 \begin{equation}
p(W_{ij} = w) = \lambda e^{-\lambda w}
\end{equation}

\noindent with $\lambda = 12.5$.

The model may present a discontinuous dynamical phase transition for large $\langle k \rangle$ and $\pi$ values, a continuous transition for intermediate values of $\langle k \rangle$ and/or $\pi$, and the absence of transitions for very low values of $\langle k \rangle$ \cite{Zarepour}. They can be characterised by the mean network activity $f_a \equiv \langle a \rangle $, where

\begin{equation}
a =  \frac{1}{N}\sum_{i=1}^N \delta(x_i,1)
\end{equation}

\noindent is the instantaneous fraction of excited neurons, and $\langle \cdots \rangle$ stands for both a temporal average and over the quenched disorder.
This quantity acts as an order parameter showing high and low activity regimes when the control parameter $T$ is varied. The fluctuation susceptibility of $f_a$ is given by

\begin{equation}
\chi = N \left(\langle a^2 \rangle -\langle a \rangle^2 \right).
\label{chi}
\end{equation}

In a second-order phase transition, this quantity is expected to present a size-dependent maximum which scales as $\max\chi \sim N^{\gamma'/\nu d}$ for $N \gg 1$, where $N$ is the number of nodes in the network, $\gamma'$ and $\nu$ are the critical exponents associated with the infinite size divergence of $\chi$ and the correlation length respectively and $d$ is the effective dimension of the model \cite{Lubeck} (namely, $d$ equals the spatial dimension if smaller than the upper critical dimension $d_c$ and $d=d_c$ otherwise). WS networks do not have a well defined dimension \cite{Newman}. The spatial dimension $D$, defined as $D=d( \log N)/d (\log l)$, where $l$ is the linear dimension of the network, behaves (for a WS network constructed from a one dimensional ring) as $D \sim \log \left(\pi\langle k \rangle N \right)$ for $\pi\langle k \rangle N \gg 1$ \cite{Newman}. Therefore, in the thermodynamic limit, they become infinite dimensional. Even in finite size systems, for the range of network sizes of our simulations, we have $D > 10$, so we can assume $d=d_c$.
 
 Finally, the first  coefficient of the autocorrelation function,
\begin{equation}
AC(1)=\frac{\langle a(t+1) -\langle a\rangle\rangle \langle a(t) -\langle a\rangle\rangle}{Var(a)},
\end{equation}
computes the average correlation of the activity, and it is useful for finding a rough estimate of $T_c$:  it is bounded ($|AC(1)|\leq 1$),  it is maximum at $T_c$, and compared to other variables such as susceptibility, it tends to have a broad peak (i.e., $AC(1)$ decays slowly with the distance to $T_c$)\cite{Chialvo2020}. In the next section, we analyse the behaviour of both $f_a$ and $\chi$ as a function of $T$ with the aid of $AC(1)$ estimates, for different system sizes $N$.

\section{Absorbing critical behaviour without spontaneous activity}
\label{r10}

When $r_1=0$, the system presents an absorbing state, namely a stationary macroscopic state where all the neurons remain quiescent for all time (except for some periodic solutions corresponding to specific initial conditions). While the system may develop a non-absorbing stationary state in the infinite size limit, for finite sizes the absorbing state is the only stationary state \cite{Marro,Henkel}. Hence, the finite size scaling of models exhibiting absorbing states is often obtained by studying the statistical properties in the quasi-stationary regime they present for several realisations of the stochastic noise between a transient and the final decay \cite{Dickman1}.  Several strategies and methods were developed to study the quasi-stationary properties, often referred to in a general way as {\it quasi-stationary methods} \cite{Dickman1,Ferreira}. Each method has advantages and disadvantages, and none is free of ambiguities. In this work, we implemented two different variations of the quasi-stationary method, which we called the {\it reactivation algorithm} and the {\it fixed time algorithm}. 

In the reactivation algorithm, the system is restarted in a new random configuration with $30\%$ of the neurons in the excited state ($x_i = 1$) every time it falls into the absorbing state. After discarding the transient period, the activity is measured again and concatenated with the previous activity. The procedure is repeated until a preset $t_{\max}$ sampling points are collected. A temporal average of the different quantities of interest is then performed. Finally, an average over disorder is made. We discard the current network after three failed reactivations in a row (defined as reactivations that fall into the absorbing state before surpassing the transient period).

In the fixed time algorithm, only networks whose temporal series of activity reach $t_{\max}$ steps without falling into the absorbing state are considered. Given a Watts-Strogatz network, independent simulations are run starting from a random initial condition in which $30\%$ of the neurons are in the excited state ($x_i =1$). Simulations are repeated for the same network until 10 such time series are obtained that reach $t_{\max}$ steps; those that fall into the absorbing state before $t_{\max}$ are discarded. These 10 time series are then averaged to produce a single representative series for that network. As in the previous algorithm, a temporal average of the quantities of interest is computed, and a disorder average is performed over different network realisations. For each network, we attempt up to 100 different initial conditions to obtain valid time series. If none of these attempts are successful, the network is discarded, and a new one is generated.
Typically, we used $t_{max} = 10^4$ for both algorithms, and at least 100 networks for the disorder average.

\begin{figure}
    \centering
    \includegraphics[width=\linewidth]{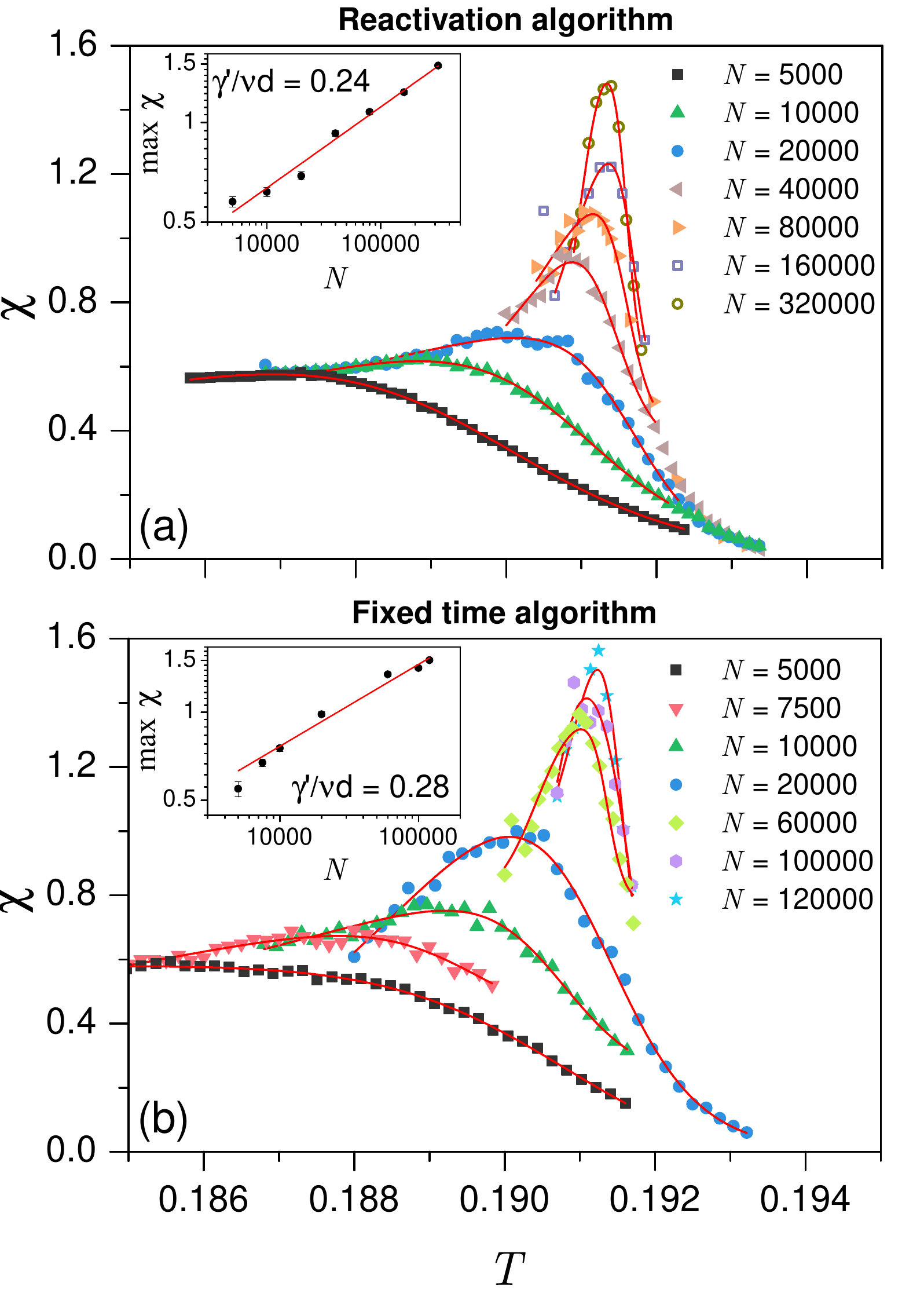}
    \caption{Susceptibility as a function of the threshold $T$ for different system sizes $N$ when $r_1=0$, obtained from different quasi-stationary algorithms. The simulations correspond to the GH model defined on a Watts-Strogatz network of size $N$ with average degree $\langle k \rangle=12$ and rewiring probability $\pi=0.6$. The continuous lines correspond to fittings using appropriate fitting functions (asymmetric double sigmoidal functions). The insets show the maxima of the different curves as a function of $N$ estimated from the fittings, with power law fittings (red lines) $\max\chi \sim N^{\gamma'/\nu d}$. The critical exponents estimations are shown in Table \ref{tab:exponents}. (a) Curves obtained using the 
 reactivation algorithm. (b) Curves obtained using the fixed time algorithm.}
    \label{fig:algoritmosr1=0}
\end{figure}

\begin{figure}
    \centering
    \includegraphics[width=\linewidth]{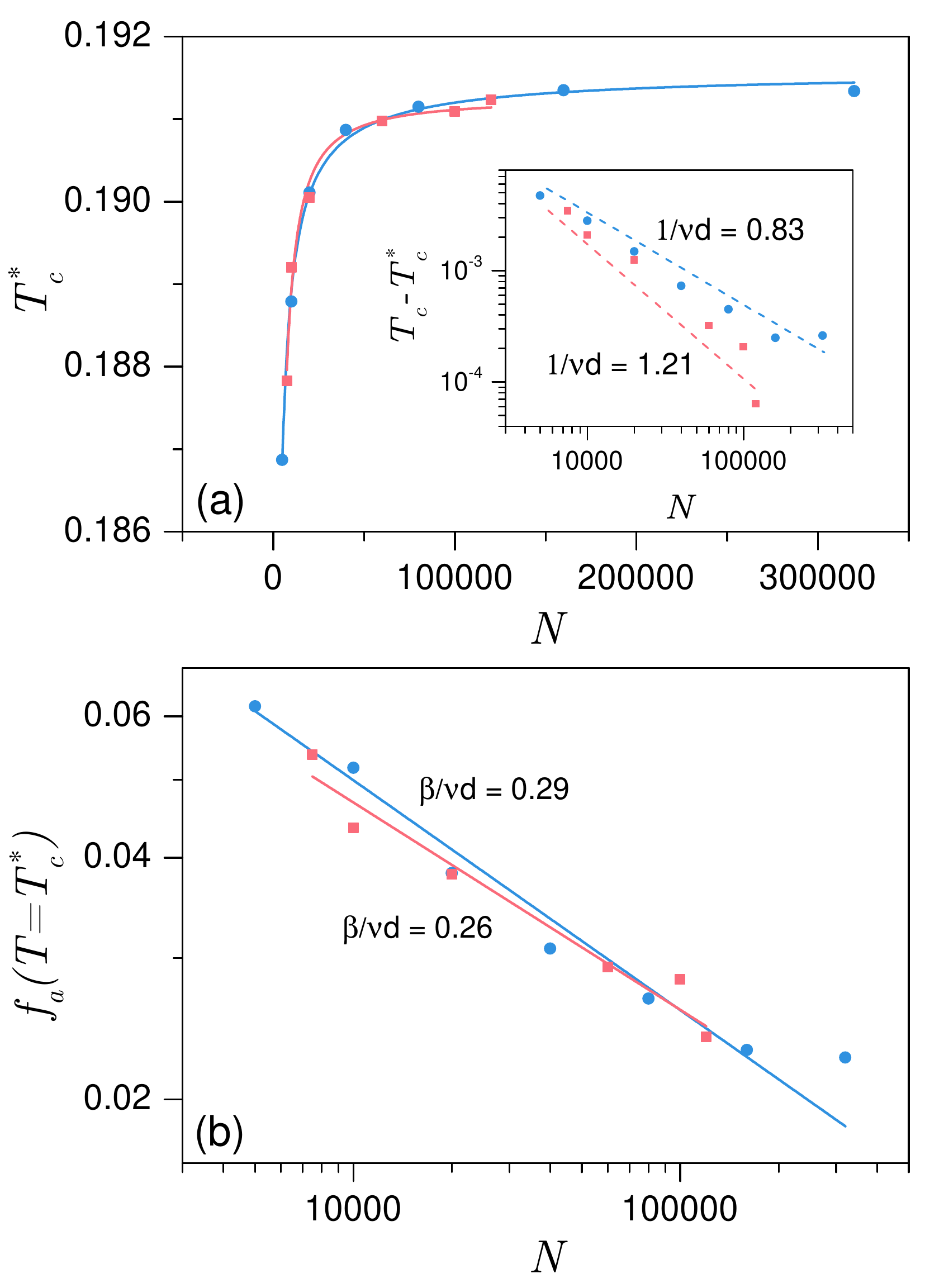}
    \caption{(a) Pseudo critical threshold $T_c^*$ estimated from the fittings of Fig.\ref{fig:algoritmosr1=0} and (b) activity $f_a$ evaluated at $T_c^*$, as a function of network size $N$ for $r_1 = 0$. Blue circles correspond to data from the reactivation algorithm and red squares to the fixed time algorithm. The continuous lines are fitting curves using  $T^*_c \sim T_c - N^{-1/\nu d}$  and $f_a(T=T^*_c) \sim N^{-\beta/\nu d}$. The estimated values of $T_c$, $1/\nu d$ and $\beta/\nu d$ from the fittings are shown in Table \ref{tab:exponents}. Inset: $T_c-T_c^*$ in log-log scale. Dashed lines are the corresponding power law fittings.}
    \label{fig:algoritmos-Tc}
\end{figure}

\vspace{1cm}

We run simulations of the GH model defined on a Watts-Strogatz network of size $N$ with average degree $\langle k \rangle=12$ and rewiring probability $\pi=0.6$, from which we obtained both the average network activity $f_a$ and the susceptibility $\chi$ from Eq.(\ref{chi}) using both algorithms for $r_1=0$. The obtained susceptibility as a function of the threshold $T$ for different system sizes is shown in Fig.\ref{fig:algoritmosr1=0} ($f_a$ as a function of $T$ will be analysed later, see Figs.\ref{fig:aprox1} and \ref{fig:aprox2}). We verified the presence of a size divergent peak of $\chi$. All the curves were fitted using appropriate fitting functions, from which the maxima of the curves were estimated as well as their location (pseudo critical points) $T_c^*(N)$. Using these values, we calculated the order parameter (activity) at the pseudo critical point $f_a(T=T_c^*)$. The results for $T_c^*$ and $f_a(T=T_c^*)$ as a function of $N$ are shown in Fig.\ref{fig:algoritmos-Tc}. We verified the expected finite size scaling behaviours in a critical phenomenon \cite{Lubeck}: $\max\chi \sim N^{\gamma'/\nu d}$, $T^*_c \sim T_c - N^{-1/\nu d}$  and $f_a(T=T^*_c) \sim N^{-\beta/\nu d}$, where $\nu = \nu_\perp$ in conventional notation, and $d$ is the effective dimension. The estimated values of $T_c$ and the critical exponents are summarised in Table \ref{tab:exponents}.

\begin{figure}
    \centering
    \includegraphics[width=\linewidth]{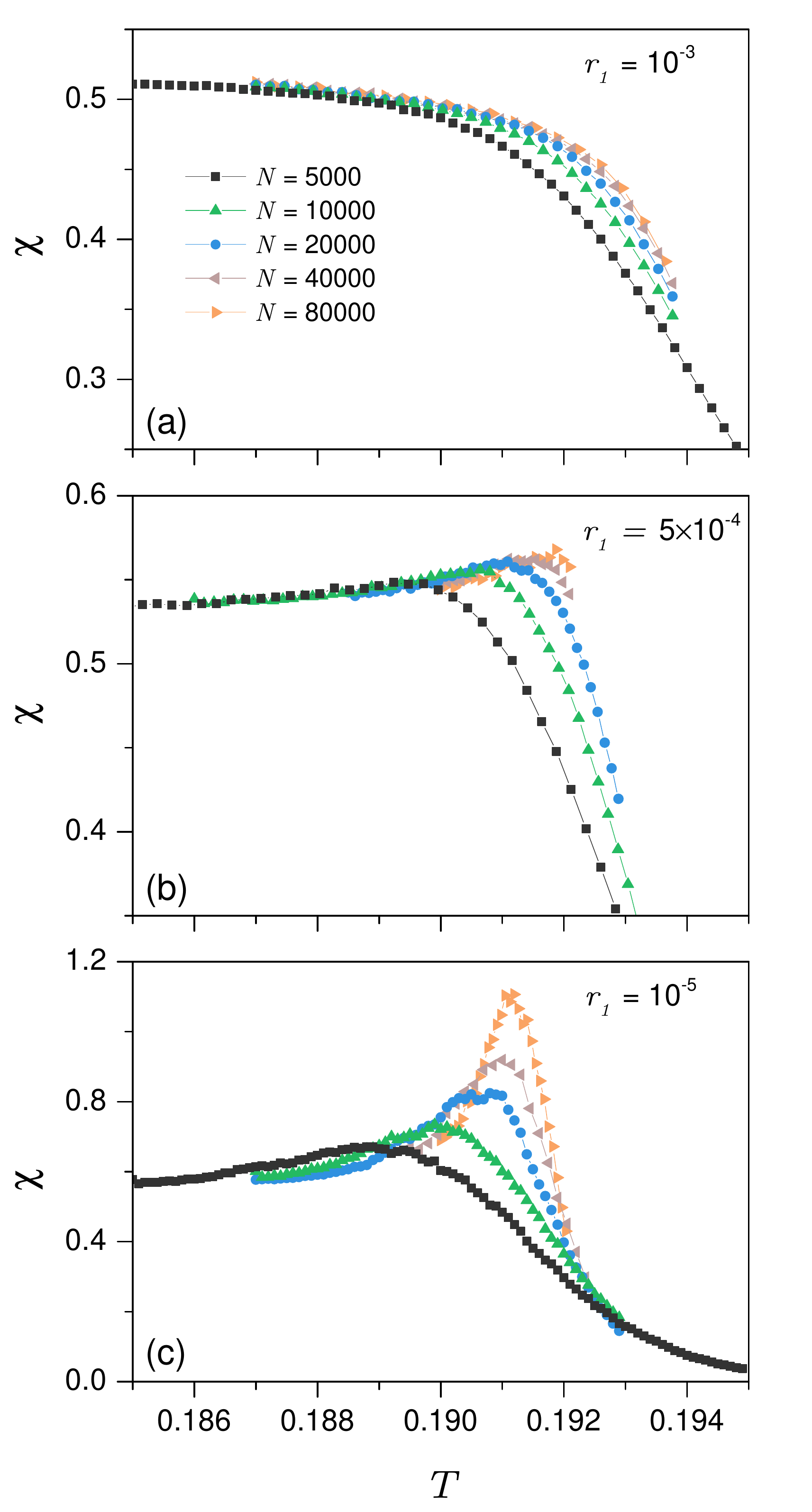}\\
    \caption{Susceptibility as function of the threshold $T$ in Watts-Strogatz networks with $\langle k \rangle=12$ and $\pi=0.6$ for different network sizes $N$ and three typical values of the spontaneous activation probability: (a) $r_1 = 10^{-3}$, (b) $r_1 = 5\times10^{-4}$ and (c) $r_1 = 10^{-5}$. The continuous lines are a guide to the eye.}
    \label{fig:suscep-nonull-r1}
\end{figure}

\begin{figure}
    \centering
    \includegraphics[width=\linewidth]{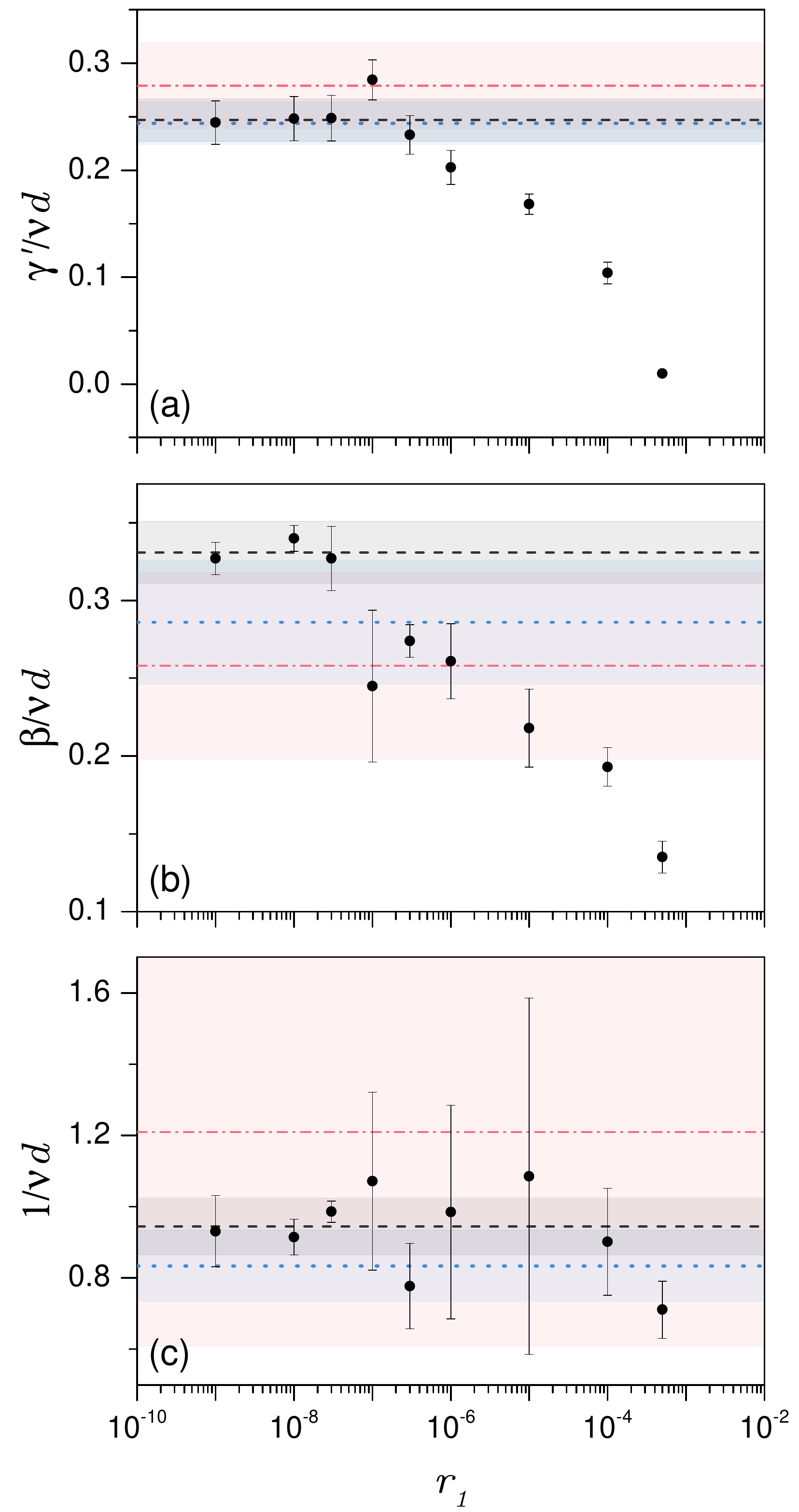}\\
    \caption{Critical exponents as a function of the spontaneous activation probability $r_1$ estimated through finite-size scaling data fittings of (a) susceptibility ($\sim N^{\gamma'/\nu d}$), (b) activity $f_a$ at $T_c^*$ ($\sim N^{-\beta/\nu d}$), and (c) pseudo critical point of the maximum of susceptibility/variance ($\sim T_c - N^{-1/\nu d}$). Examples of data used to extract these exponents are shown in Fig.\ref{fig:suscep-nonull-r1}. Black dashed lines correspond to the mean values (for the first three points) (see Table \ref{tab:exponents}).  Blue dot line and red dashed-dot line correspond to the critical exponent values obtained with the reactivation and fixed time algorithms, respectively (see Table \ref{tab:exponents}). Shadows indicate the corresponding uncertainties.}
    \label{fig:exponentes}
\end{figure}

Finally, the third method consists of running standard simulations for decreasing values of $r_1$, considering the $r_1\to 0$ limit.
In Fig.\ref{fig:suscep-nonull-r1} we illustrate the typical behaviour of $\chi$ as a function of $T$ for different system sizes for three typical values of $r_1$. While for relatively large values of $r_1$ we recover the absence of scaling \cite{Grigera}, as $r_1$ decreases, we see a clear emergence of a divergent peak with $N$. Following the same procedure as with the previous two methods, we estimated the critical exponents $\gamma'/\nu d$, $\beta/\nu d$ and $1/\nu d$, for different values of $r_1$. The results are shown in Fig.\ref{fig:exponentes}. We have a clear convergence when $r_1 \to 0$  of all the exponents to values that are similar to the ones obtained by the two $r_1=0$ algorithms. The average values when $r_1 \to 0$ are compared with those estimations in Table \ref{tab:exponents}. We see that the three estimations are consistent, giving approximate values $\beta/\nu d \approx 0.3$ and $\gamma'/\nu d \approx 0.25$ which, to the best of our knowledge, do not agree with any previously known universality class. 

Critical phenomena on systems with an absorbing state exhibit the so-called generalized hyperscaling relation \cite{Henkel,Munoz}

\begin{equation}
  \gamma' + \beta +  \beta' =\nu d
    \label{hypergen}
\end{equation}
where $\beta'$ is the exponent related to the percolation probability \cite{Henkel}. In some systems (particularly those belonging to the  Directed Percolation (DP) universality class) $\beta=\beta'$  and Eq.(\ref{hypergen}) reduces to the usual hyperscaling relation  $\gamma'+2\beta=d\nu$ \cite{Henkel,Lubeck}. This is due to the so-called rapidity reversal symmetry, which means that $f_a(t) \propto P_{sur}(t)$, where $P_{sur}(t)$ is the survival probability of finding at least one active site. In the Supplemental Material \cite{SuppMat}, we show evidence that, after a transient period, $f_a(t) \propto P_{sur}(t)$. In Table \ref{tab:exponents} we show the numerical value of $\gamma'/(\nu d) + 2\beta/(\nu d)$, obtained with the three different methods. These results suggest that $\gamma'+2\beta< d\nu$, within the confidence intervals of our statistical error bars. However, systematic errors associated with the numerical methods, as well as possible difficulties to reach the large system size asymptotic regime, may introduce a bias, so the fulfillment of the hyperscaling relation  $\gamma'+2\beta=d\nu$ (as suggested by the supplemental material results) cannot be excluded.

Finally, we analysed the behaviour of the order parameter $f_a$ as a function of $r_1$ when $r_1\to 0$, for values of $T$ close to $T_c^*$. The numerical results are shown in the Supplemental Material \cite{SuppMat}. Our results are consistent with a power law behaviour $f_a \sim r_1^{1/\delta_h}$, not only at $T=T_c^*$, but also for a wide range of values of $T$ above $T_c^*$. The exponent $\delta_h \geq 1$ depends on $T$, converging to $\delta_h=1$ as $T$ increases to values far away from $T=T_c^*$. This behaviour suggests the presence of rare--regions effects \cite{Vojta2006,Moretti}.

Summarising, the whole scenario results are consistent with the fact that $r_1$ acting as an external field for the absorbing transition is the reason behind the lack of scaling previously observed \cite{Grigera} (see Fig. \ref{fig:suscep-nonull-r1}).

\begin{table}[]
\begin{tabular}{c|ccc}
       & Reactivation & Fixed Time & $r_1\to 0$   \\ \hline
$T_c$           & 0.1916$\pm$0.0002 & 0.1913$\pm$0.0004 & 0.1918$\pm$0.0004  \\
$1/\nu d$       & 0.83$\pm$0.1 & 1.21$\pm$0.6 & 0.94$\pm$0.08  \\
$\beta/\nu d$   & 0.29$\pm$0.04 & 0.26$\pm$0.06 & 0.33$\pm$0.02  \\
$\gamma'/\nu d$ & 0.24$\pm$0.02 & 0.28 $\pm$ 0.04 & 0.25 $\pm$ 0.02 \\
$\frac{\gamma'}{\nu d} + 2\frac{\beta}{\nu d}$ & 0.8 $\pm$ 0.1  & 0.8 $\pm$ 0.2 & 0.91 $\pm$ 0.06
\end{tabular}
\caption{Summary of all exponents and critical thresholds obtained with the three methods. Uncertainties are calculated in the usual manner from fitting curves or averages where they correspond. Two standard deviations were taken as the error estimate.}
\label{tab:exponents}
\end{table}

\section{Activation mechanisms}
\label{activation}

\begin{figure}
    \centering
    \includegraphics[width=\linewidth]{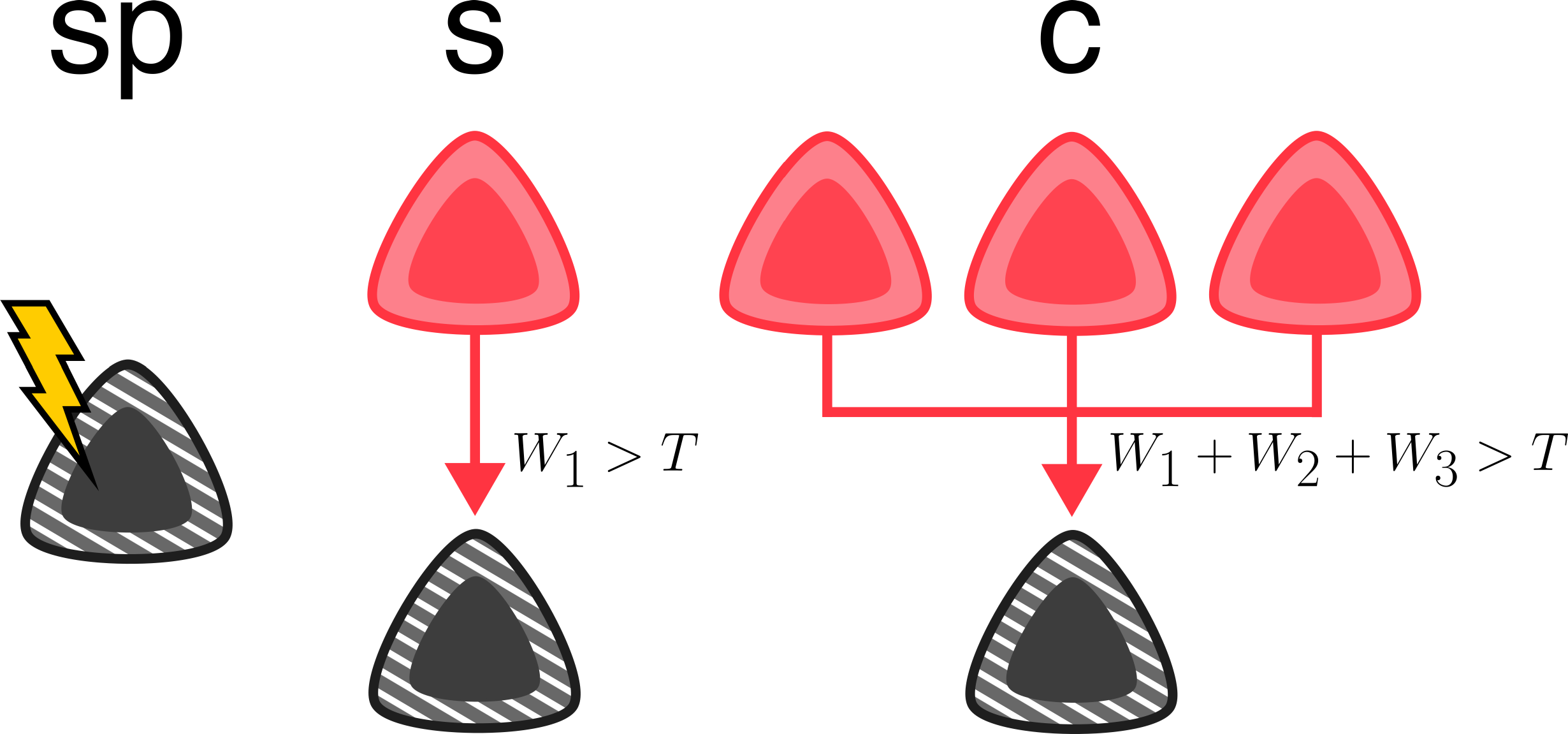}
    \caption{ Schematic drawing of the three mechanisms through which a quiescent neuron can be activated: \textit{spontaneous (sp)}, \textit{single activation (s)}, and \textit{cooperative activation (c)}. Red triangles are excited neurons, and black ones are quiescent.}
    \label{fig:dibujo}
\end{figure}

We now consider the effect of spontaneous activation and its relation to other activation mechanisms. Let us consider a quiescent neuron $i$ at time $t$. Three different mechanisms (represented in Fig.\ref{fig:dibujo}) can be identified through which it can be activated at time $t+1$: {\it (i)} \emph{spontaneous activation} (with probability $r_1$); {\it (ii)}  \emph{single activation}, that is, $i$ has one or several active neighbours at time $t$, but the strongest connection among them is enough to activate it, ($max_j(W_{i,j} \delta_{x_{j}(t),1})>T$) and {\it (iii)} \emph{cooperative activation}, where neuron $i$ is activated through the sum of the contribution of two or more active neurons ($max_j(W_{i,j} \delta_{x_{j}(t),1})<T<\sum_jW_{i,j} \delta_{x_{j}(t),1}$). The fraction of activations of each kind will be labelled   $a^{(sp)}$,  $a^{(s)}$ and $a^{(c)}$, respectively. 
In the $T\to\infty$ limit, the only possible activation mechanism is spontaneous activation. In that case, $f_a \simeq r_1$, and the net effect of the spontaneous activation is equivalent to raising the fraction of active neurons by $r_1$.

Let us consider the case for large but finite $T$, where activity is still low. Consider an excited neuron $i$ at time $t$. It may happen that  $i$ has some quiescent neighbour neuron $j$ such that $W_{i,j}>T$. In such a case, neuron $j$ will be activated (by a single activation) at the next time step, as a consequence of the activation of $i$. 
We may derive a mean-field approximation for the fraction of activity in the stationary regime at high $T$, that considers only spontaneous and single activation,  as follows. 

The fraction of active neurons at the next time step will be equal to the fraction of quiescent neurons times the probability of activation by any mechanism.

The fraction of quiescent neurons equals one minus the fraction of active neurons $a$, minus the fraction of refractory neurons. Since the refractory time lasts $1/r_2$ time steps on average, assuming stationarity, we may write the fraction of quiescent neurons as $q(t)\simeq1-(1+1/r_2) a(t)$. 

The activation probability can be written as one minus the chances of not being activated by any of the two possible mechanisms: either \emph{spontaneous activation} $(1-r_1)$, or a network activation $p_n(t)$. That is, $a(t+1)=q(t) \times [1-(1-r_1) \cdot (1-p_n(t))]$.

Assuming that each neuron interacts with the average activity on the network, when $T$ is high enough, we can approximate $1-p_n(t)\simeq [1- a(t) P(W>T)]^k$, where  $1- a(t) P(W>T)$ is the probability of not being activated by a single active neighbour, and $k$ is the number of neighbour neurons ({\it i.e.}, we do not consider here the possibility of a \emph{cooperative activation}).

Averaging and neglecting correlations (mean field approximation), we finally arrive at:

\begin{align}
f_a=&[1-(1+1/r_2) f_a] \label{aprox} \\
&\times [1-(1-r_1) \cdot (1-f_a P(W>T))^{k}]\notag.
\end{align}

For low $f_a$, we may approximate further

\begin{align}
f_a=&[1-(1+1/r_2) f_a] \label{aprox2} \\
&\times [1-(1-r_1) \cdot (1-k f_a P(W>T))]\notag.
\end{align}
\noindent so we have a degree 2 polynomial in $f_a$. 

Some remarks about Eq.(\ref{aprox2}) are in order.
First, the equation is a mean-field approximation, where the activation of a given neuron is determined by the average activity on the whole network. It does not consider the locality of neighbours (the value of rewiring $\pi$ is not considered in the equation), and consequently, it is expected to give better results for large values $\pi \to 1$.  
\begin{figure}
    \centering
    \includegraphics[width=\linewidth]{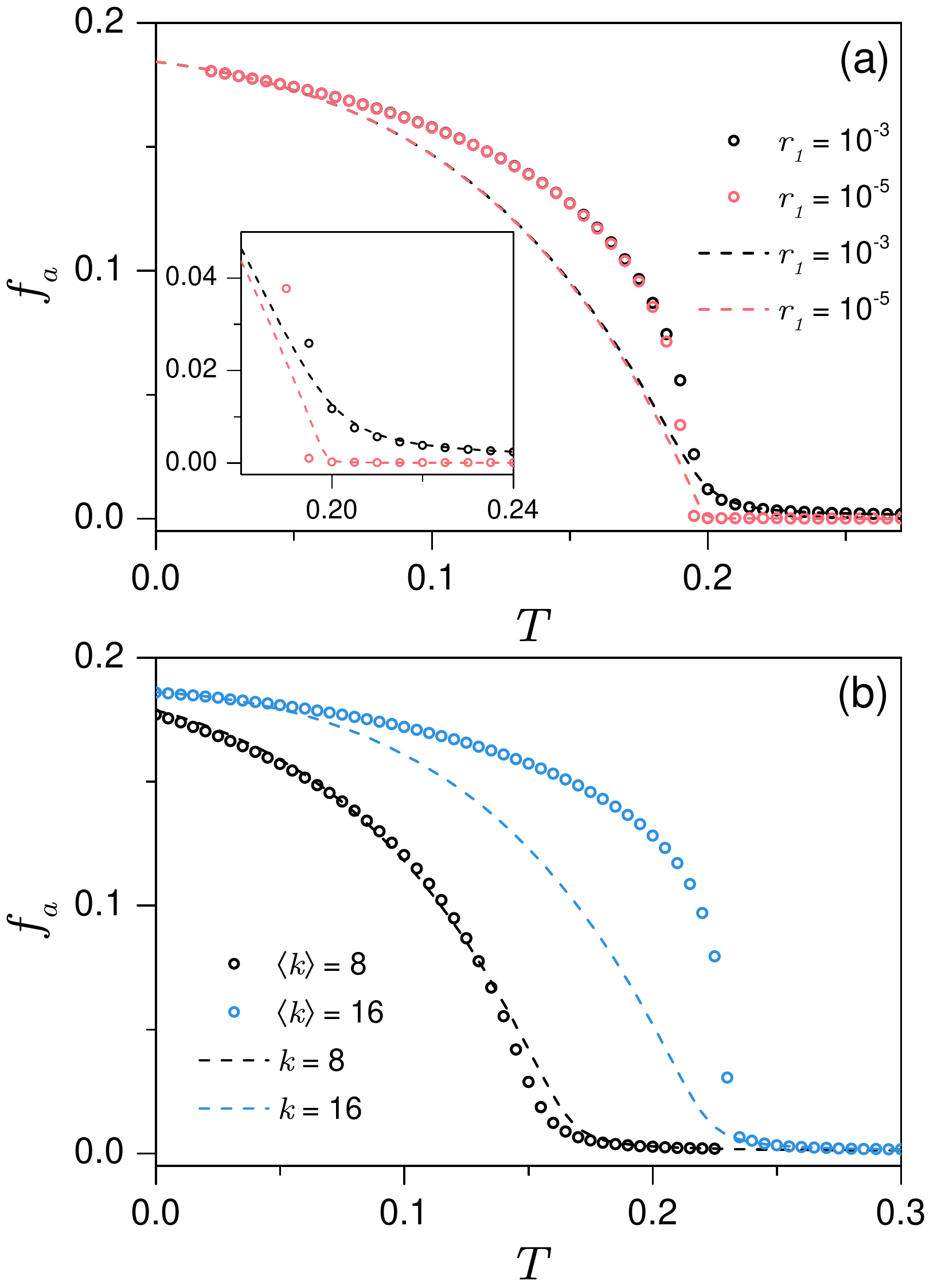}
    \caption{Average activity $f_a$ as a function of $T$ for $\pi=1$: comparison between solutions of Eq.(\ref{aprox}) (dashed lines)  and numerical simulations (symbols) for $N=10000$. (a) Different values of $r_1$ for $\langle k \rangle=12$: $r_1=10^{-3}$ (black) and $r_1=10^{-5}$ (red). (b) Different values of $\langle k \rangle$ for $r_1=10^{-3}$: $\langle k \rangle=8$ (black) and $\langle k \rangle=16$ (blue).}
    \label{fig:aprox1}
\end{figure}

Similarly, the equation accounts for random spontaneous activation and also for activation due to a single active presynaptic neuron (single activation, $a^{(s)}$). \emph{Cooperative activation} due to the contribution of several presynaptic contributions $a^{(c)}$ is not considered there. In that sense, notice that the equation is quite similar to that derived in Ref.\cite{Kinouchi}  for a different model which does not include cooperative activation at all. It is easy to see that Eq.(\ref{aprox2}) predicts a critical point at $r_1=0$ and $T_c = \ln(k)/\lambda$ where,  close to the transition, we have $f_a(T,r_1=0)\sim (T-T_c)^{\beta_{MF}}$,  $\chi'(T,r_1=0)={\partial f_a\over \partial r_1}|_{r_1=0}\sim (T-T_c)^{-\gamma_{MF}}$ and $f_a(T=T_c,r_1) \sim r_1^{1/\delta_{h,MF}}$. The critical exponents are those of mean field directed percolation exponents: $\beta_{MF}=1$, $\gamma_{MF}=1$, and $\delta_{h,MF}=2$, in agreement with the results of  Ref.\cite{Kinouchi}. We also see that, consistently with the results of section \ref{r10}, the spontaneous activation probability $r_1$ behaves as an external field conjugated to the order parameter $f_a$ of the continuous phase transition.

In Fig.\ref{fig:aprox1} we show results for the approximation Eq.(\ref{aprox2}) for several values of $k$  and $r_1$ in the continuous region, together with numerical simulation results  ($\langle k \rangle$ in the WS network of the simulations is taken equal to the value of $k$ used in the model) for $N=10000$. The approximation works reasonably well for high $T$, as expected, and fails in general for intermediate $T$ values. We will show next that, for continuous transitions, the relevance of \emph{cooperative activation} (not considered in the present approximation) takes place at intermediate values of $T$ only.  For low $T$, Eq.(\ref{aprox}) yields good results again. At that point,  \emph{cooperative activation} ceases to be relevant: there are several active neurons, and it is highly likely that one of them alone will have a connection strength larger than $T$.

\begin{figure*}
    \centering
    \includegraphics[width=\linewidth]{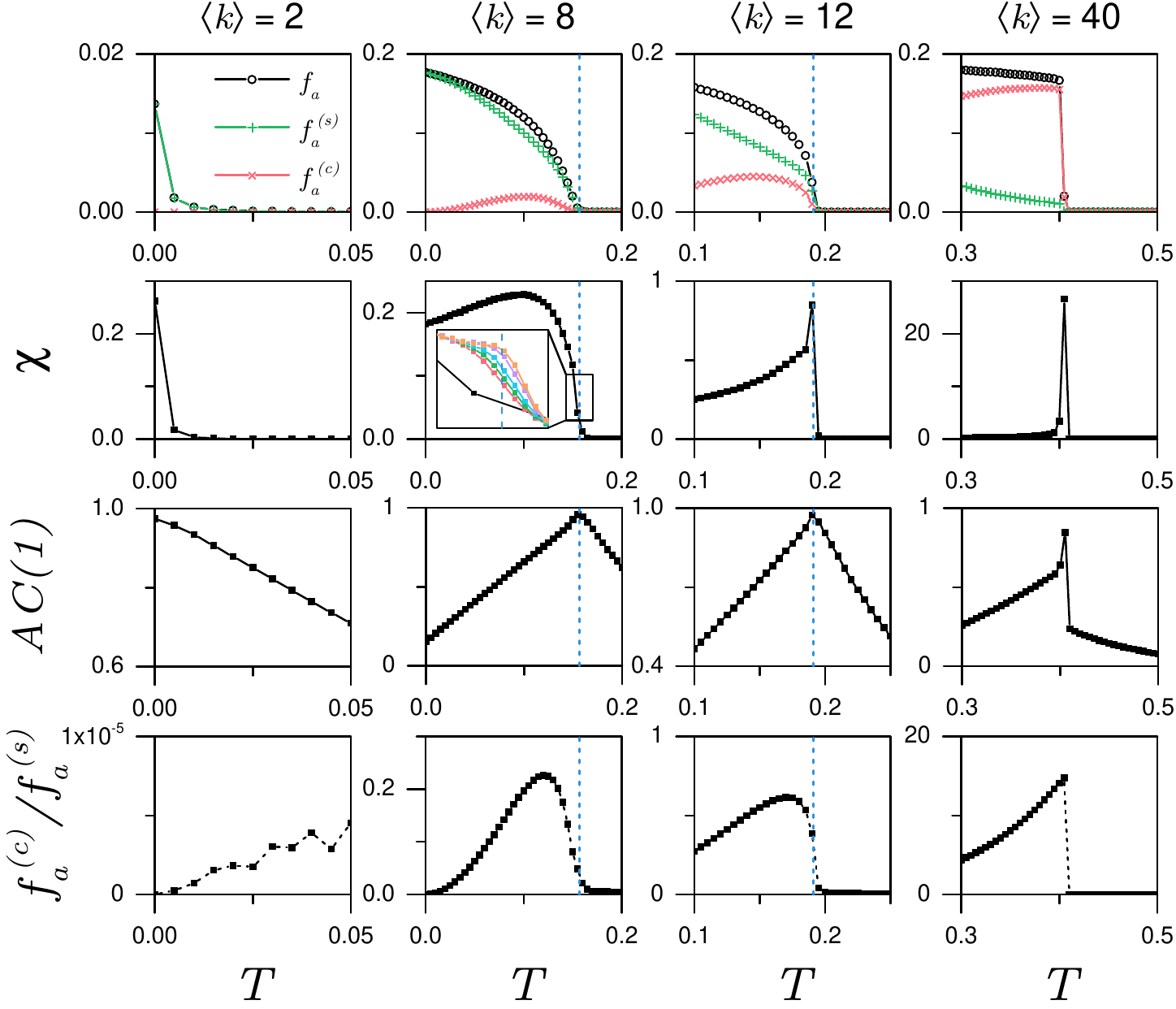}
    \caption{Activities, susceptibility  and $AC(1)$ as a function of $T$ for $r_1=10^{-5}$, $N=10000$, $\pi=0.6$ and different values of $\langle k \rangle$. Each column corresponds to one value of $\langle k \rangle$. From left to right $\langle k \rangle = 2,8,12,40$. Upper row shows the activities as a function of $T$. Intermediate rows show the susceptibility $\chi$ and the first autocorrelation coefficient $AC(1)$. Blue dashed lines mark the pseudo critical thresholds $T_c^*$ obtained from $AC(1)$ peaks. The colour curves in the inset correspond to larger system sizes ($N = 80000$ to $1280000$) and show that the scaling peak of  $\chi$ for $\langle k \rangle = 8$ only emerges for large enough network sizes. The lower row shows the ratio among \emph{cooperative} and \emph{single} contributions to the activity, $f_a^c/f_a^s$ (spontaneous activation  $f^{(sp)}=f_a-f_a^{(c)}-f_a^{(s)}$ is negligibly small and consequently not plotted).   All simulations were averaged over at least 30 networks.}
    \label{fig:aprox2}
\end{figure*}

We next performed numerical simulations to calculate the average fractions of neurons with single activation, $f_a^{(s)}\equiv \langle a^{(s)}\rangle$ and with  \emph{cooperative activation} $f_a^{(c)}\equiv \langle a^{(c)}\rangle$. The results are shown in Fig.\ref{fig:aprox2} for different values of $\langle k \rangle$, together with the total activity $f_a$ and the corresponding susceptibilities.

For very low values of $\langle k \rangle$, it is known that this system presents no phase transition of any kind \cite{Zarepour,Sanchez}, as evidenced by the monotonous decay of $f_a$ and the absence of a non-trivial peak of $\chi$ as a function of $T$. We see that in this case  ($\langle k \rangle=2$) $f_a^{(c)}$ is negligible, and the ratio $f_a^{(c)}/f_a^{(s)}$ grows monotonically with $T$. 

For intermediate values of $\langle k \rangle$ ($\langle k \rangle=8,12$ in Fig.\ref{fig:aprox2}), the system presents a second order ({\it i.e.}, continuous)  phase transition \cite{Zarepour,Sanchez}, where $\chi$ presents a non trivial peak with $T$ that scales with the system size (see section \ref{r10}). We see that $f_a^{(c)}$ and $f_a^{(c)}/f_a^{(s)}$ in this case have a continuous peak around the transition, where the transition mechanism is progressively dominated by \emph{cooperative activation} as $\langle k \rangle$ increases. Consequently,  the approximation used in Eq.(\ref{aprox2}) becomes worse (see Fig.\ref{fig:aprox1}b). Notice that for $\langle k \rangle =8$ the susceptibility presents two distinct peaks. One peak slowly grows with $N$ and it is located approximately at the same point where $AC(1)$ has its maximum. We found that the second peak does not grow with $N$ and therefore it is not associated with the phase transition. This anomaly seems to be related to the dominance of cooperative activation, since it coincides with the maximum of  $f_a^{(c)}/f_a^{(s)}$. This scenario appears for $ 4 \leq \langle k \rangle \leq 10$. For values of $ 12 \leq \langle k \rangle \leq 24$, the second peak disappears and the growing maximum becomes sharper, even for relatively low values of $N$.

Finally, for large values of $\langle k \rangle$ ($\langle k \rangle=40$ in Fig.\ref{fig:aprox2}) the transition becomes first order ({\it i.e.}, discontinuous) \cite{Zarepour,Sanchez}, hysteresis appears and both the order parameter $f_a$ and the susceptibility $\chi$ present a discontinuous jump, without scaling with the system size (not shown). In this case, we see that the activation mechanism becomes almost completely cooperative. This result is consistent with that shown in Ref.\cite{Sanchez}, where another related model, without cooperative activation at all, presents no discontinuous phase transition.



\begin{figure}
    \centering
    \includegraphics[width=\linewidth]{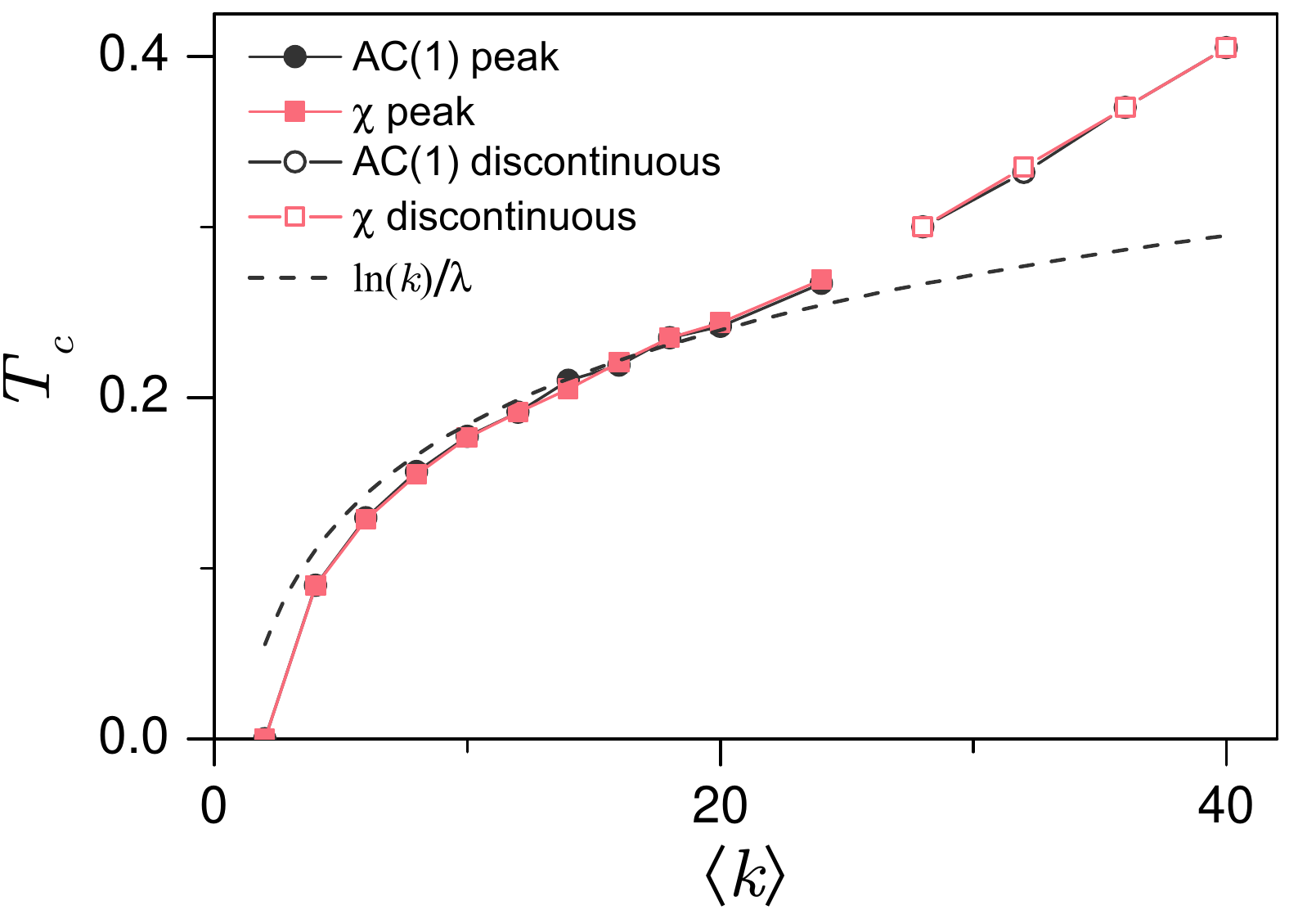}
    \caption{Critical threshold $T_c$ for $r_1=0$ as a function of the average degree $\langle k \rangle$ in the continuous regime $4 \leq \langle k \rangle \leq 24$ and metastability limit in the discontinuous regime $\langle k \rangle > 24$,  for $N=10000$ and $\pi=0.6$. Estimates were computed as the values that maximise $\chi$  and the first autocorrelation coefficient $AC(1)$. The dashed line corresponds to the mean field approximation prediction.}
    \label{fig:aprox3}
\end{figure}

Another way of estimating the critical point value $T_c$ is through the first autocorrelation coefficient $AC(1)$ of the order parameter fluctuations, which presents a maximum at $T_c$ in continuous phase transitions \cite{Chialvo2020} ($AC(1) \to 1$ when $N\to\infty$ at the maximum) and it also has a peak at the metastability limits in discontinuous phase transitions \cite{Sanchez}.


In Fig.\ref{fig:aprox3} we compare the estimation of $T_c$ from the maximum of $AC(1)$  and the non- trivial ({\it i.e.}, that scales with $N$) peak of the susceptibility as a function of $\langle k \rangle$ for $N=10000$ and $r_1=0$, using the reactivation algorithm (see section \ref{r10}). Checks were made for some values of $\langle k \rangle$ by extrapolating the $N\to\infty$ limit of $T_c^*$ in the continuous regime and also with standard simulations with $N=10000$ and $r_1=10^{-5}$. The differences were negligible in all the cases.
We see that the estimations obtained from both quantities ($AC(1)$ and $\chi$) are indistinguishable for all values of $\langle k \rangle$. However the peak of $AC(1)$ is broad (i.e., it decays slowly as $|T-T_c|$ is increased) while the peak of $\chi$ may be narrow for some network topologies. Because of that, a reasonable estimate of $T_c$ can be found from $AC(1)$, for all studied values of $\langle k \rangle$, even for small values of $N$.

In the discontinuous regime ($\langle k \rangle > 24$), we find that $T_c$ grows proportional to $\langle k \rangle$, which is expected since  the relevance of \emph{cooperative activation} is maximum there, and the number of active neighbours should
be proportional to $\langle k \rangle$. 

In the continuous regime ($4 \leq \langle k \rangle \leq 24$), the estimations agree very well with our mean field 
approximation prediction $T_c=\ln(k)/\lambda$, almost in the whole range of values of $\langle k \rangle$, consistently with the fact that \emph{cooperative activation} is small close to the critical point for continuous transitions.

\section{Discussion}
\label{Discussion}

We analysed the finite size scaling of the fluctuation susceptibility $\chi$ of the GH model on a Watts-Strogatz network with $r_1=0$, using two different quasi-stationary methods, as well as standard numerical simulations in the limit $r_1\to 0$. While most of our analysis focused on a particular network topology $\langle k\rangle =12$ and $\pi=0.6$, some checks for other topology parameter values in the continuous phase transition regime range gave consistent results.

We found that the susceptibility exhibits the finite size scaling behaviour expected for a critical point when $r_1=0$. In particular, the size dependent peak in the susceptibility that emerges when $r_1 \to 0$ coincides with the maximum of the first autocorrelation coefficient for large enough system sizes. For large enough values of $r_1$, the susceptibility divergence disappears, which explains the lack of scaling reported in Ref.\cite{Grigera}. The whole phenomenology suggests that $r_1$ plays the role in this model of an external field conjugated to the order parameter $f_a$. 

The usage of three different calculation methods allowed us to obtain reliable estimations of the critical exponents, which do not seem to agree (to the best of our knowledge)  with any known universality class. The last fact is probably associated with the presence of quenched disorder in the model, since rare regions effects \cite{Vojta2006,Moretti} in this case cannot be discarded (a possibility also supported by the apparent presence of power law behaviours for many different values of $T$ shown in the supplementary material); future works along this line will include critical exponents calculations varying the disorder, either in the synapsis probability distribution or in the WS rewiring probability. Another extension of the present work could be to perform a finite size scaling analysis in the $r_1 \to 0$ limit of the variability in the lasting time activity when starting from a single neuron activation, as done in Ref.\cite{Haimovici} for a fixed system size.

We also analysed the influence of different activation mechanisms around the activation transition point by comparing numerical simulations with a mean field approximation that neglects cooperative activations. We found that cooperative mechanisms become progressively relevant as the average degree of the network increases, being almost negligible for small values of $\langle k \rangle$ and dominant for large values of it. In particular, in the range of values of $\langle k \rangle$ where the activation transition is continuous, those mechanisms contribute to the phenomenon only for intermediate values of the threshold $T$. Consequently, the mean field approximation works well for low and high values of $T$, including the critical region. 
In particular, the prediction of a logarithmic increase of the critical value $T_c$ with $\langle k \rangle$ when $r_1=0$, agrees very well with the numerical results in almost the whole region of continuous phase transitions $4 \leq \langle k \rangle \leq 24$. These results are consistent with those of a recent work\cite{Messe}, where an equation similar to Eq.(\ref{aprox2}) is derived under the same assumption that single activations are dominant. The validity of this assumption was supported by both theoretical and computational evidence for a wide range of threshold values in different network topologies and weight distributions. Moreover, they also show that cooperative activation becomes more relevant as $\langle k \rangle$ increases, consistently with our results.

Our mean field approximation also predicts a critical phenomenon in the DP universality class, in agreement with the prediction of a related model without cooperative interactions at all \cite{Kinouchi}. However, the critical exponents for the model defined on a WS network obtained numerically do not agree with any known universality class, particularly with the mean field DP ones that would be expected in this case. This result could be consistent with the fact that the hyperscaling relation seems not to be satisfied.  However, the numerical evidence of rapidity reversal symmetry in the model (see supplementary material) makes the whole scenario inconclusive, leaving the question about universality in the model defined on a WS network open.

The result $f_a(T=T_c,r_1) \sim r_1^{1/\delta_{h,MF}}$ within the mean field approximation reinforces the idea that $r_1$ acts as an external field conjugated to $f_a$. While the numerical results in a WS network also support this conclusion, they are consistent with a threshold dependent critical exponent $\delta_h(T)$ for a wide range of values $T\geq T_c$, thus providing another evidence of the presence of rare-regions effects.

Finally, in the region $\langle k \rangle > 24$, where the dynamical phase transition becomes discontinuous and is dominated by the cooperative activation mechanism, $T_c$ grows linearly with $\langle k \rangle$, as expected.

\section*{Data Availability}
The data supporting this study are available in Ref.~\cite{Data}

\begin{acknowledgments}
This work was partially supported by CONICET (Argentina) through Grants PIP No. 1122020010106, by SeCyT (Universidad Nacional de Córdoba, Argentina) and by the NIH (USA) Grant No.
1U19NS107464-01. J.A. supported by a Doctoral Fellowship from CONICET (Argentina). This work used Mendieta Cluster from CCAD-UNC, which is part of SNCAD-MinCyT, Argentina. DRC thanks Hong Kong Baptist University for funding his Distinguished Professorship of Science during these studies.
\end{acknowledgments}





\end{document}